\documentclass[aps,12pt]{revtex4}
\usepackage{graphicx}
\usepackage{amsmath}
\usepackage{color}

\begin{document}
\newcommand{\del}{\delta\:\!\!}
\newcommand{\amie}{m_i/m_e}
\newcommand\isg{\epsilon_g}
\newcommand\qal{q_\alpha}
\newcommand\mal{m_\alpha}
\newcommand\qmal{\frac{q_\alpha}{m_\alpha}}
\newcommand\etal{{\it et.~al.\ }}
\def\btwdl{\tilde {\bf b}}
\newcommand{\wdot}{\dot w}
\newcommand{\vdot}{\dot v_H}
\newcommand{\vdoto}{\dot v_{H0}}
\newcommand{\vdotl}{\dot v_{H1}}
\newcommand{\ldot}{\dot \lambda_H}
\newcommand{\ldotl}{\dot \lambda_{H1}}
\newcommand{\ldoto}{\dot \lambda_{H0}}
\newcommand{\vg}{{\bf V}_H}
\newcommand{\vgl}{{\bf V}_{H1}}
\newcommand{\vgo}{{\bf V}_{H0}}
\newcommand{\DDt}{\frac{D}{Dt}}
\newcommand{\pa}{{\scriptscriptstyle{\|}}}
\newcommand{\pe}{{\scriptscriptstyle{\perp}}}
\newcommand{\eps}{\varepsilon}
\newcommand{\lac}{L_{ac}}
\newcommand{\lk}{L_k}
\newcommand{\dm}{D_m}
\newcommand{\lp}{\lambda_{\pa}}
\newcommand{\lpp}{\lambda_{\pe}}
\newcommand{\lc}{\lambda_c}
\newcommand{\cpar}{\chi_{\pa}}
\newcommand{\cpp}{\chi_{\pe}}
\def\({\left (}
\def\){\right )}
\def \<{\left \langle}
\def \>{\right \rangle}
\def\[{\left [}
\def\]{\right ]}
\def\vxi{\boldsymbol{\xi}}
\def\vrho{\boldsymbol{\rho}}
\def\vkappa{\boldsymbol{\kappa}}
\def\veta{\boldsymbol{\eta}}
\def\vlp{\mathopen{\boldsymbol{(}}}
\def\vrp{\mathclose{\boldsymbol{)}}}
\arraycolsep=2.5pt
\def\nfrac#1#2{{{\displaystyle #1}\over{\displaystyle #2}}}
\def\mfrac#1#2{\displaystyle{\frac{#1}{#2}}}
\newcommand{\ddt}[1]{\frac{d #1}{dt}}
\newcommand{\dd}[2]{\frac{d #1}{d #2}}
\newcommand{\ppt}[1]{\frac{\partial #1}{\partial t}}
\newcommand{\beq}{\begin{eqnarray}}
\newcommand{\eeq}{\end{eqnarray}}
\newcommand{\beqn}{\begin{equation}}
\newcommand{\eeqn}{\end{equation}}
\newcommand{\aver}[1]{\left\langle #1 \right\rangle}
\newcommand{\nabs}[1]{\left|\!\, #1 \,\!\right|}
\newcommand{\aveg}[1]{\overline{#1}}
\newcommand{\df}{\delta  \! f}
\newcommand{\raw}{\rightarrow}
\newcommand{\vperp}{v_{\pe}}
\newcommand{\vpar}{v_{\pa}}
\newcommand{\ppar}{p_{\pa}}
\newcommand{\kpar}{k_{\pa}}
\newcommand{\kperp}{k_{\perp}}
\newcommand{\lpar}{\lambda_{\pa}}
\newcommand{\lperp}{\lambda_{\perp}}
\newcommand{\vth}{v_t}
\newcommand{\dcv}{d^3 \! {\bf v}}
\newcommand{\dcx}{d^3 \! {\bf x}}
\newcommand{\vder}[2]{\frac{\partial #1}{\partial {\bf #2}}}
\newcommand{\vdero}[1]{\frac{\partial}{\partial {\bf {#1}}}}
\newcommand{\der}[2]{\frac{\partial #1}{\partial #2}}
\newcommand{\dersq}[2]{\frac{\partial^2 #1}{\partial {#2}^2}}
\newcommand{\dero}[1]{\frac{\partial}{\partial #1}}
\newcommand{\derosq}[1]{\frac{\partial^2}{\partial {#1}^2}}
\newcommand{\dbdp}{\frac{\partial B}{\partial \psi_p}}
\newcommand{\dbdt}{\frac{\partial B}{\partial \theta}}
\newcommand{\dbdz}{\frac{\partial B}{\partial \xi}}
\newcommand{\dadp}{\frac{\partial \widetilde{\alpha}}{\partial \psi_p}}
\newcommand{\dadt}{\frac{\partial \widetilde{\alpha}}{\partial \theta}}
\newcommand{\dadz}{\frac{\partial \widetilde{\alpha}}{\partial \xi}}
\newcommand{\padt}{\frac{\partial \widetilde{\alpha}}{\partial t}}
\newcommand{\dptdp}{\frac{\partial \Phi}{\partial \psi_p}}
\newcommand{\dptdt}{\frac{\partial \Phi}{\partial \theta}}
\newcommand{\dptdz}{\frac{\partial \Phi}{\partial \xi}}
\newcommand{\pptdt}{\frac{\partial \Phi}{\partial t}}
\newcommand{\dedb}{\mu + \rho_pl^2 B}
\newcommand{\grad}{{\bf \nabla}}
\newcommand{\gperp}{{\bf \nabla}_{\pe}}
\newcommand{\gpar}{{\bf \nabla}_{\pa}}
\newcommand{\hs}{\hspace{.3in}}
\def\I{{\bf I}}
\newcommand{\bhat}{{\bf \hat{b}}}
\newcommand{\bddel}{{\bf \hat{b}}\cdot \nabla}
\newcommand{\bdotdel}{{\bf \hat{b} \cdot \nabla}}
\newcommand{\om}{\omega}
\newcommand{\oma}{\omega_A}
\newcommand{\gb}{\gamma_b}
\newcommand{\gm}{\gamma_{MHD}}
\newcommand{\bh}{\beta_h}
\newcommand{\ba}{\beta_{\alpha}}
\newcommand{\Om}{\Omega}
\newcommand{\al}{{\widetilde {\alpha}}}
\newcommand{\pfo}{\omega_{d0}}
\newcommand{\wat}{\tilde{\omega}_A}
\newcommand{\ws}{\omega_*}
\newcommand{\wis}{\omega_{*i}}
\newcommand{\whs}{\omega_{* \alpha}}
\newcommand{\abs}[1]{\mid #1 \mid}
\newcommand{\ten}[1]{\vec{\vec{#1}}}
\newcommand{\pp}{\delta {\bf P}}
\newcommand{\db}{\delta {\bf B_{\pe}}}
\newcommand{\dbp}{\delta {\bf B_{\pa}}}
\newcommand{\xip}{{\bf \xi}_{\pe}}
\newcommand{\ve}[1]{{\bf #1}}
\newcommand{\wka}{\delta W_k^{(a)}}
\newcommand{\wkn}{\delta W_k^{(na)}}
\newcommand{\wknt}{\delta W_{k,t}^{(na)}}
\newcommand{\wknp}{\delta W_{k,c}^{(na)}}
\newcommand{\dwk}{\delta W_k}
\newcommand{\bnn}{\begin{displaymath}}
\newcommand{\enn}{\end{displaymath}}
\newcommand{\xipsi}{\xi_{\psi_p}}
\newcommand{\nuef}{\nu_{\rm eff}}
\newcommand{\dt}{\triangle t}
\newcommand{\exb}{\rm E\!\times\!B}
\newcommand{\qovm}{\frac{q}{m}}
\def\marked{blue}
\newcommand{\odr}{\mathcal{O}}
\newcommand{\nn}{\nonumber}
\newcommand{\jac}{\mathcal{J}}

\title{Saturation of the kinetic ballooning instability due to the electron parallel nonlinearity}
\author{Yang \surname{Chen}}
\email{yang.chen@colorado.edu}
\affiliation{Center for Integrated Plasma Studies, University of Colorado at Boulder, Boulder, CO 80309}
\author{Scott E. \surname{Parker}}
\affiliation{Renewable and Sustainable Energy Institute, University of Colorado at Boulder, Boulder, CO 80309}

\begin{abstract}
  The electron parallel nonlinearity (EPN) is implemented in the gyrokinetic Particle-in-Cell turbulence code GEM
  [Y. Chen and S. E. Parker, J. Comp. Phys. 220, 839 (2007)]. Application to the Cyclone Base Case reveals a
  strong effect of EPN on the saturated heat transport above the kinetic ballooning mode threshold.
  Evidence is provided to show that the strong effect
  is associated with the electron radial motion due to magnetic fluttering, which turns fine structures of the KBM eigenmode in
  radius into fine structures in velocity and increases the magnitude of the EPN term in the kinetic equation.
\end{abstract}

\maketitle

Gyrokinetic simulations that are based on a gyrokinetic equation for the perturbed distribution function $\df$
often neglects the so-called parallel nonlinearity \cite{dlee}, which has the form
$\dot{v}_{\rm \pa 1} \partial\df/\partial v_\pa$ where $\dot{v}_{\rm \pa 1}$ is the particle acceleration along the magnetic field
due to the electromagnetic fluctuations. This common practice has its origin in the first formulation of nonlinear gyrokinetics,
namely, the Frieman-Chen equation \cite{frieman82}
which is derived based on the gyrokinetic ordering $\delta=\rho/L_{\rm eq}\ll 1$. Here $\rho$ is
the Larmor radius and $L_{\rm eq}$ is a scale-length characterizing the spatial variation of the plasma equilibrium.
Only terms of $\odr(\delta^2)$ are retained in the Frieman-Chen equation, and the parallel nonlinearity is of
$\odr(\delta^3)$ and negligible. Indeed,  simulations of the electrostatic
Ion-Temperature-Gradient-Driven (ITG) turbulence with adiabatic electrons were
little affected by the ion parallel nonlinearity \cite{candy2006}.

It is not clear if the electron parallel nonlinearity (EPN) can be similarly neglected in ITG turbulence, and if its role changes
in the presence of fluctuating magnetic fields. 
Here we describe the implementation of EPN in the $\df$ Particle-in-Cell code GEM \cite{chen03a,chen07a} and report numerical observations
on EPN.

We provide details of GEM only as needed for describing the impact of EPN on the algorithm.
GEM uses the parallel canonical momentum as the parallel velocity coordinate. However, in order to indicate the appearance of all
EPN-related terms, it is convenient to start with the electron drift-kinetic equation in the $v_\pa$-formulation,  
\beqn
\frac{\partial \df}{\partial t} + \mathbf{v}_G \cdot \grad \df+\dot{v}_\pa\frac{\partial \df}{\partial v_\pa} =
   - \mathbf{v}_{\rm G1} \cdot \grad f_M - \dot{\eps}_k  \frac{\partial f_M}{\partial \eps_k}.
\label{eq:gkdf}
\eeqn
The distribution function is decomposed into the equilibrium distribution $f_0$, often
approximated with a Maxwellian $f_M$, and the perturbation $\df$; $v_\pa$ is the velocity
parallel to the equilibrium magnetic field ${\bf B}_0$, $\dot{v}_\pa$ is the parallel acceleration; $\eps_k$ is defined below,
and ${\bf v}_G$ is the guiding-center velocity,
\beqn
{\bf v}_G=v_\pa\frac{{\bf B}^*}{B_\pa^*}+\frac{\mu}{qB_\pa^*}{\bf b}\times \grad B_0 + \frac{1}{B_\pa^*}{\bf b}\times {\grad \phi}
   \equiv {\bf v}_{\rm G0}+{\bf v}_{\rm G1}
\eeqn
with
\beqn
{\bf B}^*={\bf B}_0+\frac{m}{q}v_\pa\grad\times {\bf b}+\db
\eeqn
\beqn
B_\pa^*=B_0+\frac{mv_\pa}{q}{\bf b}\cdot\grad\times {\bf b}
\eeqn
\beqn
    {\bf v}_{\rm G1}=v_\pa\frac{\db}{B_\pa^*} + \frac{1}{B_\pa^*}{\bf b}\times \grad \phi
\eeqn    
Here ${\bf b}={\bf B}_0/B_0$ is the unit vector along ${\bf B}_0$, $\mu=mv_\pe^2/2B_0$ is the magnetic moment,
$m=m_e$ is the electron mass, and $q=-e$ is the electron charge.
The electromagnetic field fluctuation 
 is given by $\delta {\bf E}=-\grad\phi-\ppt{A_\pa}{\bf b}$ 
and  $\db=\grad A_\pa \times {\bf b}$. 
The equilibrium
distribution is assumed to be Maxwellian
\beqn
f_M({\bf R},\eps_k)=\frac{n({\bf R})}{(2\pi)^{3/2}v_T^3}\exp{\(-\eps_k/T({\bf R})\)},
\eeqn
where $\eps_k=\mu B_0+\frac{1}{2}mv_\pa^2$ is the particle kinetic energy, $T({\bf R})$ is the temperature,
$n({\bf R})$ is the density, $mv_T^2=T({\bf R})$ defines the thermal velocity $v_T$, and ${\bf R}$ is the
guiding-center position.

\newcommand{\bstar}{{\bf b}^*}
\newcommand{\invb}{\frac{1}{B_\pa^*}}
\newcommand{\bstarz}{{\bf b}_0^*}
\def\pzd0{\dot{v}_{\rm \pa 0}}

Define
\beqn
\bstarz=\frac{1}{B_\pa^*}\left ({\bf B}_0+\frac{m}{q}v_\pa\grad\times{\bf b}\right ).
\eeqn
The parallel acceleration is given by \cite{bh07}
\beqn
\dot{v}_\pa = -\frac{\mu}{m}\bstarz\cdot\grad B_0 -\frac{\mu}{m}\frac{\delta{\bf B}_\pe}{B_\pa^*}\cdot\grad B_0 
-\frac{q}{m}\bstarz\cdot \grad\phi - \frac{q}{m}\invb \delta{\bf B}_\pe\cdot \grad\phi-\frac{q}{m} \ppt{A_\pa},
\label{eq:vpadot}
\eeqn
and the rate of change of the kinetic energy is
\beqn
\dot{\eps}_k = -q{\bf v}_G\cdot\grad\phi-qv_\pa\ppt{A_\pa}.
\eeqn
A partially linearized simulation is a simulation in which the last four terms on the
right-side of Eq.~\ref{eq:vpadot} are
neglected. These terms are $\odr(\delta^2)$ while the first term is $\odr(\delta)$.  

To include EPN it is necessary to transform from $v_\pa$ to the parallel canonical momentum
\beqn
p_\pa=v_\pa+\frac{q}{m}A_\pa \label{eq:canonical}
\eeqn
as the velocity coordinate, with the goal of eliminating $\partial A_\pa/\partial t$ from the equations.
To this end it is also necessary to replace $\eps_k$ with
\beqn
\eps_p=\mu B_0+\frac{1}{2}m p_\pa^2,
\eeqn
and to understand the Maxwellian accordingly as $f_M({\bf R},\eps_p)$. The distinction between $f_M({\bf R},\eps_k)$
and $f_M({\bf R},\eps_p)$ is important, and will be indicated explicitly henceforth by writing either $f_M(\eps_k)$ or $f_M(\eps_p)$.

Define $\dot{v}_{\rm \pa 0}=-\frac{\mu}{m}\bstarz\cdot\grad B_0$, the rate of change of
the new variables are
\beq
\dot{p}_\pa&=& \dot{v}_{\rm \pa 0}  + \qovm{\bf v}_{\rm G0}\cdot\grad A_\pa - \frac{1}{m}\frac{\delta{\bf B}_\pe}{B_\pa^*}\cdot\mu\grad B_0 
-\frac{1}{m}\bstarz\cdot q\grad\phi \label{eq:ppadot} \\
&\equiv&\dot{v}_{\rm \pa 0}+\dot{p}_{\rm \pa 1} \nn
\eeq
and
\beqn
\dot{\eps}_p = -q{\bf v}_{\rm G0}\cdot\grad\phi + qv_\pa {\bf v}_{\rm G0}\cdot\grad A_\pa+qA_\pa \dot{v}_{\rm \pa 0}
+ qA_\pa\dot{p}_{\rm \pa 1}. \label{eq:epdot}
\eeqn
Let's examine the magnitude of $\dot{p}_{\rm \pa 1}$ in Eq.~\ref{eq:ppadot} according to the ion-scale gyrokinetic ordering,
\beq
&&\frac{e\phi}{T}\sim\frac{ev_{\rm Ti} A_\pa}{T}\sim\frac{k_\pa}{k_\pe}\sim\frac{\omega}{\Omega_i}\sim\frac{\rho_i}{L_{\rm eq}}
\sim\delta \ll 1, \\
&&k_\pe\rho_i\sim 1,
\eeq
with $v_{\rm Ti}$ being the ion thermal speed.
The ordering $e\phi/T\sim ev_{\rm Ti}A_\pa/T$ overestimates the magnitude of the magnetic fluctuation. In fact, for nearly
all drift-waves the inequality $ev_{\rm Ti}A_\pa \ll e\phi$ is satisfied. A possible exception
is the micro-tearing mode for which the electric potential is not essential. In any case, the magnitude of $A_\pa$ can be assumed to
satisfy $ev_{\rm Te}A_\pa/T_e \ll 1$, even though the electron thermal speed $v_{\rm Te}$
is much larger than the ion thermal speed. This
additional ordering implies $\dot{p}_{\rm \pa 1}\ll \dot{v}_{\rm \pa 0}$, and the terms in Eq.~\ref{eq:ppadot} and Eq.~\ref{eq:epdot}
containing $\dot{p}_{\rm \pa 1}$ should be negligible, even for electrons. However, fine structures of $\df$ in $v_\pa$,
if nonlinearly generated, will increase the magnitude of EPN relative to the $E\times B$ nonlinearity and make it potentially
important.
This point will be discussed after we see
the simulation results.

In the $p_\pa$-formulation the decomposition of the total distribution becomes $f=f_M(\eps_p)+\df_{p_\pa}$.
The equation for $\df_{p_\pa}$ is 
\beqn
\frac{\partial \df_{p_\pa}}{\partial t} + \mathbf{v}_G \cdot \grad \df_{p_\pa}+\dot{p}_\pa\frac{\partial \df_{p_\pa}}{\partial p_\pa} =
   - \mathbf{v}_{\rm G1} \cdot \grad f_M(\eps_p) - \dot{\eps}_p  \frac{\partial f_M(\eps_p)}{\partial \eps_p}.
\label{eq:gk}
\eeqn
This equation is not obtained by strictly applying the coordinate transformation to Eq.~\ref{eq:gkdf}. A term of
$\odr(\delta^3)$ that is linear in $A_\pa$ and proportional to the density/temperature gradient is neglected.

The implementation of $\dot{p}_{\rm \pa 1}$ in the equation of motion, Eq.~\ref{eq:ppadot}, and in the electron weight equation
(via $\dot{\eps}_p$) is straightforward \cite{mishchenko2021,hager2022}. These terms have been previously implemented in GEM.
However, the previous implementation is incomplete.
As a particle code, GEM uses the split-weight scheme to enhance the time step and improve numerical stability.
An essential component of the split-weight scheme is 
the vorticity equation for the rate of change of $\phi$ \cite{chen07a}. The previous implementation of EPN neglects
high order corrections due to EPN in the vorticity equation. For this reason, most previous GEM simulations are partially linearized. 

The rate of change of the electron density $\dot{\delta n_e}$
enters the vorticity
equation as a source term, and as a result the velocity space integral of the last term in Eq.~\ref{eq:gk} needs to be calculated.
In particular, part of $-\dot{\delta n_e}$ comes from the two middle terms in the expression of $\dot{\eps}_p$ in Eq.~\ref{eq:epdot},
\beqn
I=-\int v_\pa \frac{1}{T}f_M(\eps_p) {\bf v}_{\rm G0} \cdot\grad A_\pa \,2\pi\jac d\mu dv_\pa
-\int\frac{1}{T}A_\pa \dot{v}_{\rm \pa 0}f_M(\eps_p) \,2\pi\jac d\mu dv_\pa.
\eeqn
Here $\jac$ is the velocity space Jacobian defined by $d{\bf v}=\jac d\mu dv_\pa d\xi$, $\xi$ being the gyro-angle.
The two integrals in $I$ must be combined into a divergence form to avoid grid-scale numerical instabilities \cite{chen07a},
\beqn
I=I_p+\int\frac{ev_\pa}{T^2} A_\pa^2 \dot{v}_{\rm \pa 0} f_M(\eps_k)  \,2\pi \jac d\mu dv_\pa
+\int \frac{ev_\pa^2}{T^2} A_\pa \({\bf v}_{\rm G0} \cdot\grad A_\pa\) f_M(\eps_k) \,2\pi \jac d\mu dv_\pa, \label{eq:div}
\eeqn
with
\beqn
I_p\equiv -\grad\cdot\int v_\pa {\bf v}_{\rm G0} A_\pa  \frac{1}{T}f_M(\eps_p) 2\pi\jac d\mu dv_\pa. \label{eq:ip}
\eeqn
In deriving Eq.~\ref{eq:div} the incompressibility of the guiding-center motion and the
equilibrium kinetic equation have been used \cite{chen07a}.
The appearance of $f_M(\eps_k)$ in the last two terms of Eq.~\ref{eq:div} arises from an
approximate relation between the two Maxwellians,
\beqn
f_M(\eps_p) \approx f_M(\eps_k)+\frac{e}{T}v_\pa A_\pa f_M(\eps_k).
\label{eq:frelation}
\eeqn
The quantity $I_p$  is in the required divergence form. The velocity integral
is discretized using the discrete marker particles \cite{chen07a}. 

To summarize the EPN implementation, the $\dot{p}_{\rm \pa 1}$ term in Eq.~\ref{eq:ppadot} and Eq.~\ref{eq:epdot}, and
the last two terms in Eq.~\ref{eq:div}  are all
implemented. 
A partially linearized simulation corresponds to the neglect of these terms, together with
the distinction between $p_\pa$ and $v_\pa$, i.e. neglecting the second term on the right side of Eq.~\ref{eq:canonical}.

We now study the effects of EPN by comparing partially linearized simulations with simulations including EPN.
We use a global form of the Cyclone Base Case (CBC) \cite{gorler2016pop,dimits2000}.
The plasma has concentric circular flux surfaces,
a single deuterium ion species and electrons with a mass ratio of $m_i/m_e=3600$.
The major radius is
$R_0=1.67m$, the aspect ratio is $a/R=0.36$. The on-axis toroidal magnetic field is $B_0=2T$, the safety factor profile
  is $q(r)=2.52 (r/a)^2-0.16 r/a+0.86$ with $q=q_0=1.41$ at $r/a=0.5$.
  The temperature at $r_0/a=0.5$ is $T_e=T_i=T_0=2.14$ keV,
  the normalized ion Larmor radius is $\rho^*\equiv\sqrt{m_iT_i}/(eB_0a)=0.00556$. The density profile is given by
\beqn
n(r)=n_0\exp\[-\kappa_n w_n\frac{a}{R_0}\tanh\(\frac{r-r_0}{aw_n}\)\],
\eeqn
and the temperature profile is
\beqn
T(r)=T_0\exp\[-\kappa_T w_T\frac{a}{R_0}\tanh\(\frac{r-r_0}{aw_T}\)\],
\eeqn
with $\kappa_T=6.96$, $\kappa_n=2.23$, $w_T=w_n=0.3$. 
 The intensity of 
 electromagnetic effects is adjusted by scaling the density profile to give the desired value of the total plasma beta
 at $r=r_0$,  $\beta=4\mu_0 n_0T_0/B_0^2$. 
Parallel magnetic perturbation ($\delta B_\pa$) is neglected, and
there is no equilibrium flow. A pitch-angle scattering collision model \cite{chen07a} is applied to the electrons.
The electron collision rate
$\nu_e=n_0^4e^4\ln\Lambda Z_{\rm eff}/4\pi\epsilon_0^2m_e^2v_{\rm Te}^3$ 
can be estimated if $Z_{\rm eff}$ is known. Here we somewhat arbitrarily choose a fixed collision rate of $\nu_ea/c_s=0.36$
($c_s=\sqrt{T_0/m_i}$ being the ion sound speed at $r_0$), corresponding to $\beta=0.04$, $Z_{\rm eff}=1$ and $\ln\Lambda=20.8$.
A heat source with a rate of $\gamma_H a/c_s=0.036$ is applied
to both ions and electrons \cite{chen2023}.
We focus on the mode with the toroidal mode number $n=20$, corresponding to a binormal wavenumber of $k_y\rho_i=0.31$. 
Here $\rho_i=m_i v_{\rm Ti}/q_iB$ is the ion Larmor radius at $r=r_0$ and $v_{\rm Ti}=\sqrt{T_0/m_i}=c_s$ is
the ion thermal speed.
Linear simulations scanning over $\beta$ show the
familiar transition from ITG at low beta to KBM for $\beta\geq 0.03$. 

Simulation of electromagnetic turbulence for CBC has long been known to be difficult \cite{chen03b}. For moderate beta values 
well below the KBM threshold the simulation either fails to saturate or saturates at
very high fluctuation amplitudes \cite{waltz2010pop}. This phenomenon has been called the high-$\beta$ runaway \cite{waltz2010pop} or
nonzonal transition (NZT) \cite{pueshel2013}. The phenomenon is believed to be due to the suppression of
the secondary zonal flow instabilities
in the presence of magnetic stochasticity \cite{terry2015,nevins2011}. The heat fluxes from nonlinear simulations of KBM turbulence
for experimental plasmas typically much exceed the experimentally inferred fluxes, sometimes by orders of magnitudes \cite{kumar2021}.
A validation study of KBM on DIII-D uses linear simulations to obtain the KBM mode characteristics,
resorting to quasilinear theory for an estimate of the KBM-induced transport \cite{jian2023}. 
Owing to its importance to the understanding of anomalous transport, the numerical study of runaway/NZT
has been broadened in various directions, including the use of alternative electron models
\cite{dong2019gtc,chen2023gtc}, initializing the simulation with zonal flows \cite{rath2022},
including nonlocal physics such as entropy advection in the radial direction \cite{masui2022,ishizawa2024},
and the effects of EPN \cite{mishchenko2021,mishchenko2022}. Of particular importance to the present work is the study by
Mishchenko \etal \cite{mishchenko2021} which specifically discusses
 EPN and reports KBM simulations that saturate at a moderate level. 

\begin{figure}[!htp]
\includegraphics[width=0.6\textwidth]{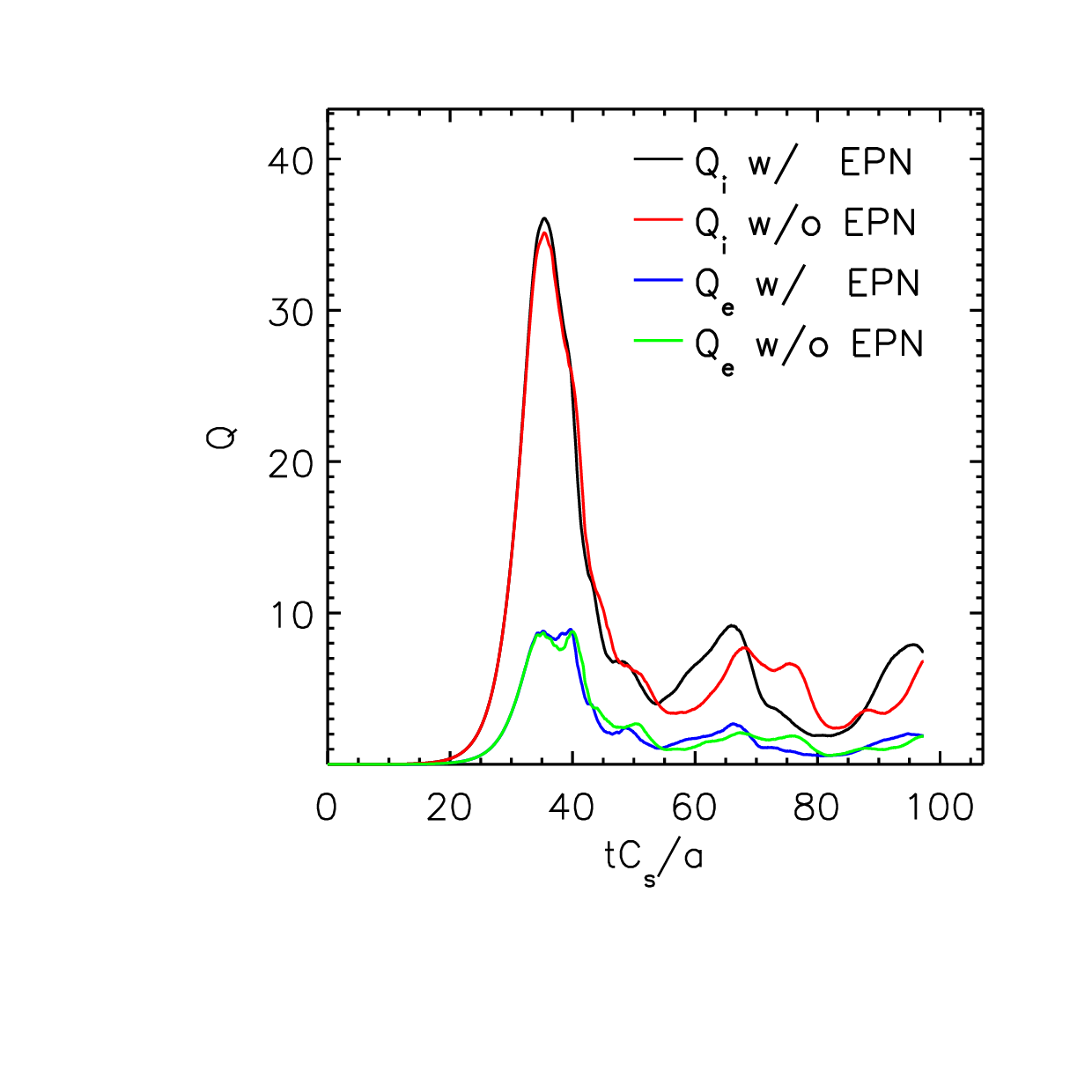} 
\caption{Time history of the ion and electron heat flux in electrostatic ITG simulation with $\beta=4\times 10^{-4}$.}
\label{fig:w}
\end{figure}

We first simplify the problem by including only toroidal modes
with $n=0$, $20$, $40$,
$60$ and $80$ in the fields $(A_\pa, \phi)$. A more comprehensive simulation is presented later. 
The simplified problem excludes all modes except a few harmonics of
$n=20$ that can be generated through the self-coupling of $n=20$. 
The high-n toroidal harmonics are very weakly unstable; their primary role in the simulation is to broaden the mode spectrum
allowed by self-coupling of the primary $n=20$ mode. 
The nonlinear mechanisms being addressed extend upon the drift-wave/zonal-flow paradigm that focuses the
interaction between a single $n>0$  drift-wave and the $n=0$ mode \cite{dong2019gtc,chen2023gtc}.
It is important to note that no filtering is applied to the distribution function.
The simulation domain is $r/a\in [0.2,0.8]$, with a grid setting of
$(n_x,n_y,n_\pa)=(1024,64,64)$, the number of grids in the radial, binormal and parallel direction, respectively.
Two cases of $\beta$ are considered, a nearly electrostatic case with $\beta=4\times 10^{-4}$ and a strongly
electromagnetic case with $\beta=0.04$. The time step
is $\triangle t c_s/a=0.00139$ for the former and $\triangle t c_s/a=6.95\times 10^{-4}$ for the latter.
The marker number is $32/{\rm cell}$ for ions and $64/{\rm cell}$ for electrons. Simulations
are initialized with ion weights set to small values and electron weights set to zero. 
We first check the EPN effects on electrostatic simulations.
Fig.~\ref{fig:w} shows the evolution of the heat fluxes for $\beta=4\times 10^{-4}$, from both the
partially linearized simulation and the simulation with EPN. All heat fluxes are the average value in $r/a\in [0.3,0.7]$, and expressed
in unit of the gyro-Bohm heat flux $Q_{\rm GB}=n_0T_0 v_{\rm Ti}(\rho_i/a)^2$. Since our focus is the EPN as a saturation
mechanism, we are only interested in the early phase of the nonlinear stage. As is
typical of electrostatic simulations, the unstable mode grows to a sufficiently large amplitude so that zonal flows are
simultaneously excited and suppress the ITG growth, leading to saturation. Fig.~\ref{fig:w} indicates that this nonlinear mechanism
is not altered substantially by EPN. This is consistent with the previous study of the ion parallel nonlinearity
in electrostatic ITG turbulence \cite{candy2006}.
It is still remarkable, though, that EPN has little effect on the initial saturation at high amplitude, despite the large
ion-to-electron mass ratio.

\begin{figure}[!htp]
\includegraphics[width=0.6\textwidth]{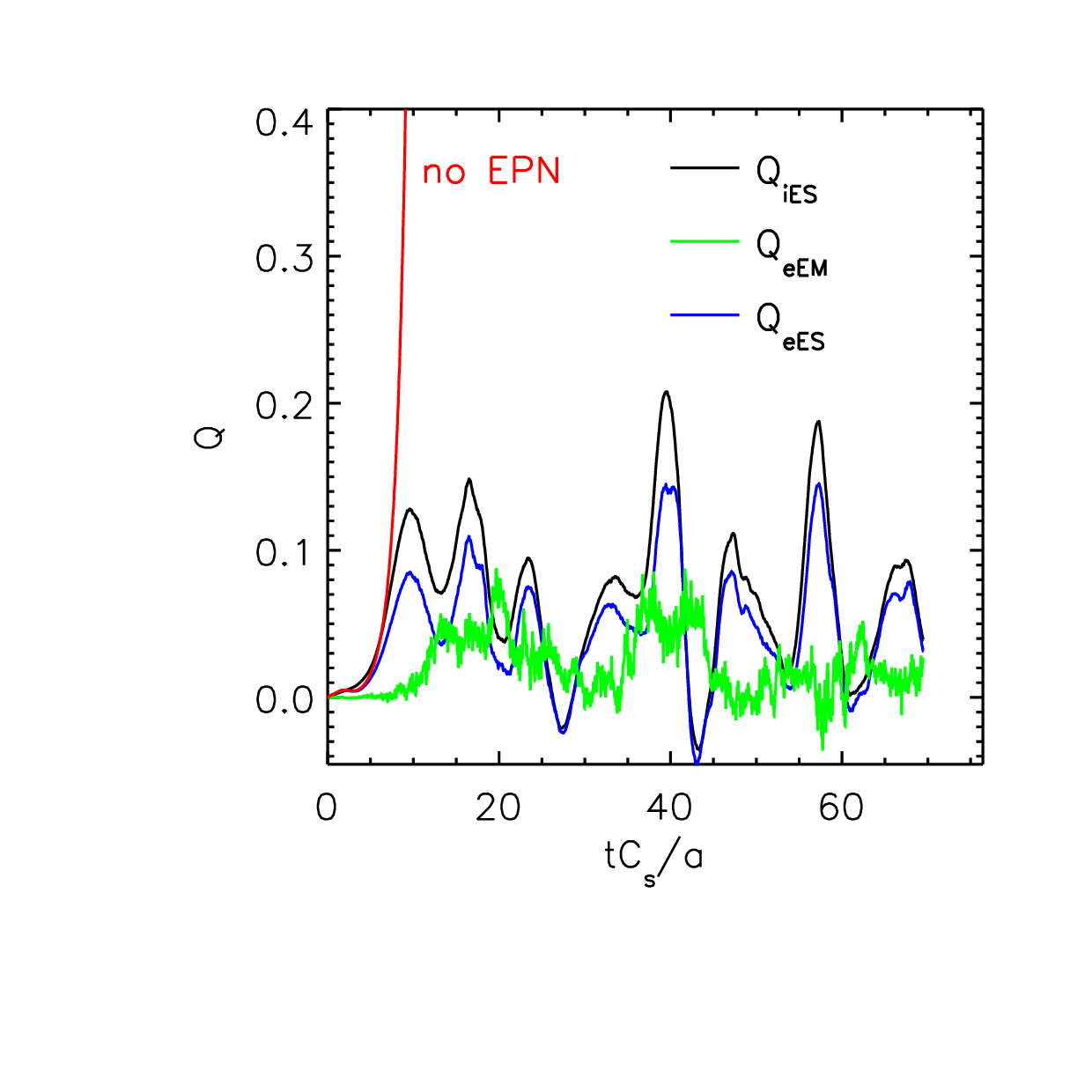} 
\caption{Time history of heat fluxes for $\beta=0.04$ with KBM. $Q_{\rm iES}$ is the $E\times B$ contribution to the ion
  heat flux, $Q_{\rm eEM}$ is the magnetic fluttering contribution to the
electron heat flux. The simulation without EPN shows runaway (red).}
\label{fig:fluxcmpx}
\end{figure}

\begin{figure}[!htp]
\includegraphics[width=0.6\textwidth]{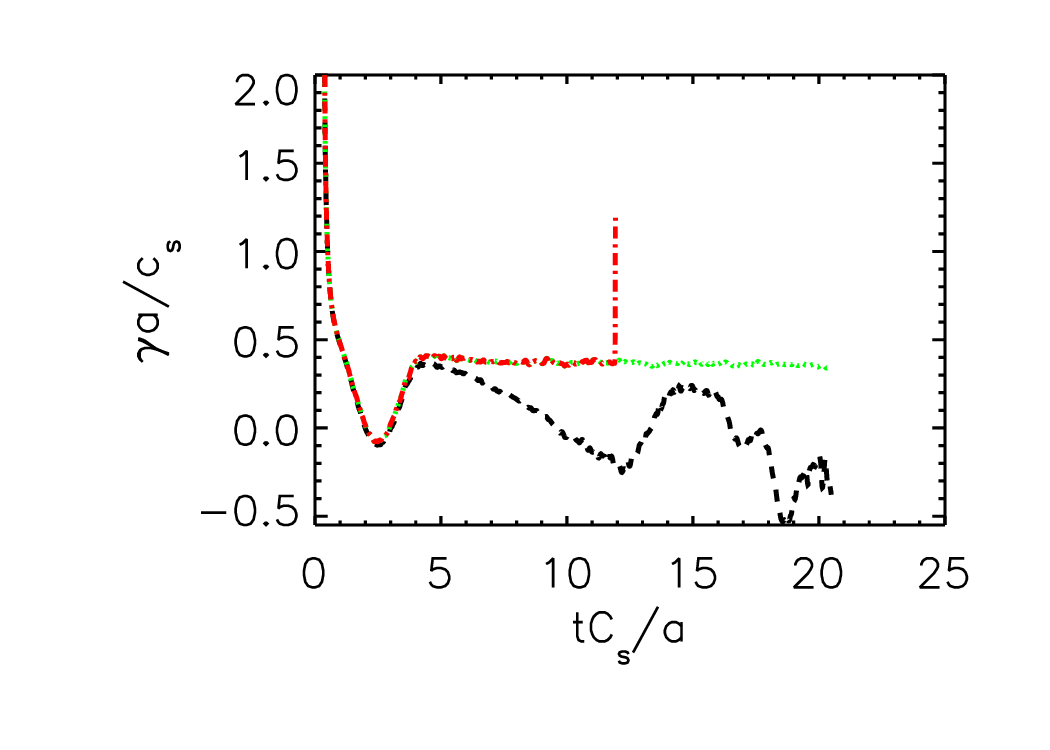} 
\caption{Time history of instantaneous growth rate in fully nonlinear simulation (black), fully linearized simulation (green) and
  partially linearized simulation (red).}
\label{fig:gamcmpx}
\end{figure}

The nonlinear evolution is drastically changed in the high-$\beta$ regime. 
The linear KBM threshold for $n=20$ is about $\beta\sim 0.03$.
Linear simulation of $n=20$ at $\beta=0.04$ yields a KBM with real frequency of $\omega_r a/c_s=0.94$
and a growth rate of $\gamma_{\rm lin}a/c_s=0.34$.
Fig.~\ref{fig:fluxcmpx} shows the evolution of the heat fluxes. This simulation will be repeated later varying the
electron mass to both well below and well above the physical value, hence the time step is reduced by half to ensure uniform
numerical stability. We first observe that 
runaway occurs  early in the nonlinear phase of the partially linearized simulation.
Because no filtering or hyper-viscosity is applied to the
particle distribution, runaway appears to be more severe in a particle simulation than in Eulerian simulations.  
The simulation with
EPN, on the other hand, saturates at an initial amplitude that is two orders of magnitude below that of
the electrostatic case.
The evolution of the fluxes with EPN deviates from the exponential growth at a fluctuation amplitude of $\phi$
much lower than typically seen with electrostatic ITG. Fig.~\ref{fig:gamcmpx} compares the instantaneous
growth rate in the nonlinear simulation with a strictly linearized simulation. Deviation from the
exponential growth occurs at $t c_s/a\approx 5$ with a volume-averaged (in the sense of root-mean-square) magnitude of
$e \nabs{\phi}/T_0=0.0013$.
By comparison, deviation occurs at $t c_s/a\approx 30$ with $e\nabs{\phi}/T_0=0.019$ in the simulation of Fig.~\ref{fig:w}.  
The evolution of the growth rate for the partially linearized simulation is also shown in Fig.~\ref{fig:gamcmpx},
with a much longer phase of linear growth prior to the onset of runaway.  

We now propose an explanation of the unusually strong EPN effects. The magnitude of the EPN term in Eq.~1 is proportional to
$\partial \df/\partial v_\pa$. EPN can be important only if, as a result of the
$E\times B$ drift and magnetic fluttering, fine structures in $v_\pa$ have developed. 
The explanation is based on the fact
that fine radial structures are
always present in the electron distribution function, and the fact
that magnetic fluttering 
 enhances the radial mixing of the distribution. These facts imply early onset of nonlinear effects.
We start with an observation on the relative importance of the $E\times B$ drift and the drift due to magnetic fluttering.
For the $n=20$ KBM the ratio between the volume-averaged magnitude of the magnetic fluctuation and
that of the electrostatic fluctuation in
the fully linear simulation is $v_u\nabs{A_\pa}/\nabs{\phi}=0.14$.
Here $v_u=\sqrt{T_0/m_p}$ is the thermal speed of a proton used for normalization.
 This ratio is still small, but for electrons
the ratio between the magnitude of the radial velocity due to magnetic fluttering and that due to $E\times B$ is    
$v_{\rm Te}\nabs{A_\pa}/\nabs{\phi}=5.9$. The radial displacement of electrons is dominated by magnetic fluttering.

\begin{figure}[!htp]
\includegraphics[width=0.6\textwidth]{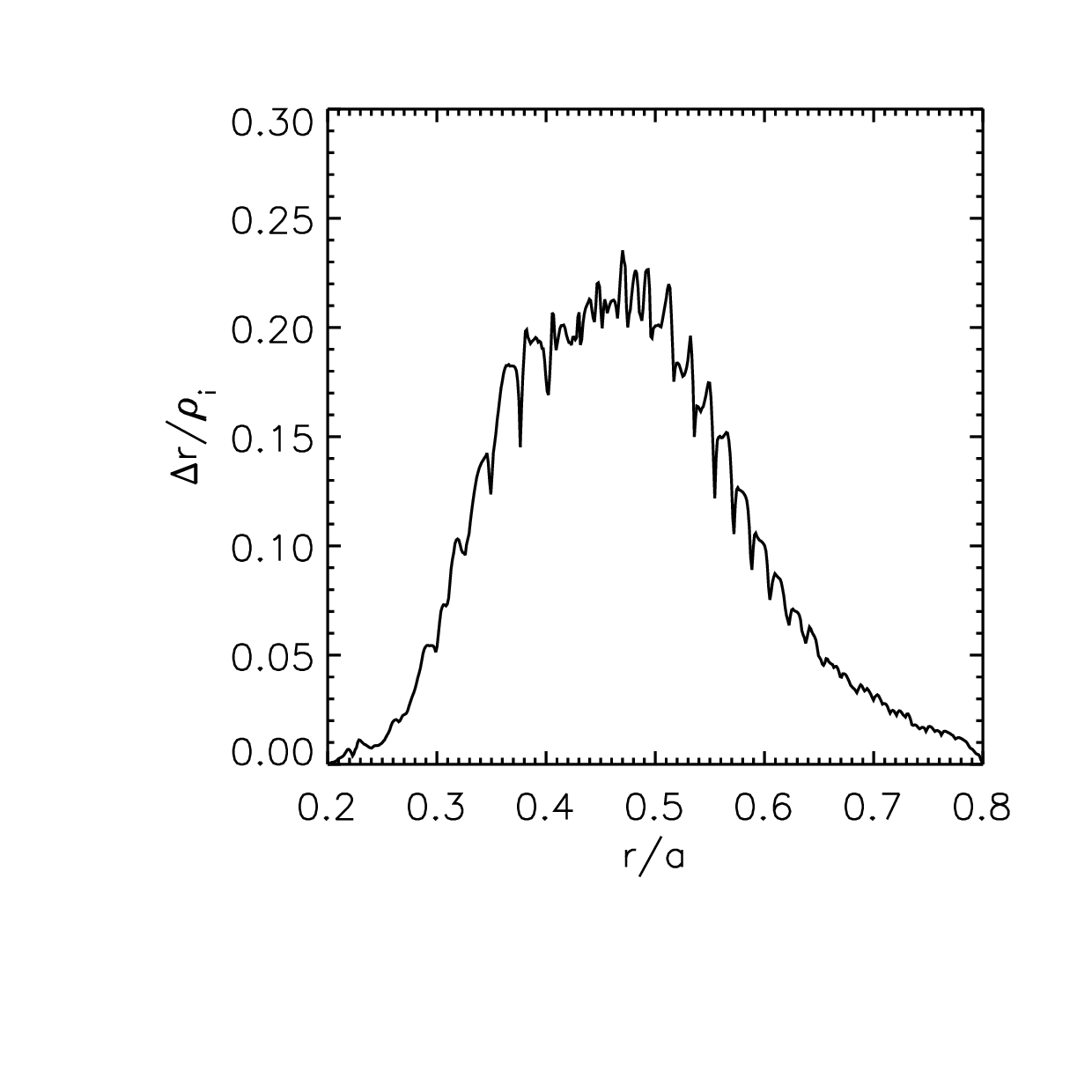}
\caption{Maximum radial excursion of the magnetic field line vs. radius.}
\label{fig:displ}
\end{figure}

 The radial displacement of passing electrons 
can be measured by the radial excursion of the magnetic field line in a poloidal
revolution. The field line excursion in $x$ and $y$, denoted by $\triangle x(\theta)$
and $\triangle y(\theta)$, 
is obtained by integrating the equations for the field line,
\beq
\frac{d\triangle x}{d\theta}=\frac{\delta{\bf B}_\pe\cdot\grad x}{{\bf B}_0\cdot\grad\theta} \\
\frac{d\triangle y}{d\theta}=\frac{\delta{\bf B}_\pe\cdot\grad y}{{\bf B}_0\cdot\grad\theta},
\eeq
and the radial width of the field line can be computed as
${\rm max}_{\theta\in[-\pi,\pi]}\[\triangle x(\theta)\]-{\rm min}\[\triangle x(\theta)\]$.
This has been done for each of the field line labeled by an $x-y$ grid at $\theta=-\pi$ (the high field side),
using  the magnetic perturbation data at $tc_s/a=5$.  Fig.~\ref{fig:displ} shows the maximum field line width on a flux surface, $\triangle r$,
as a function of radius. At the mode location, $\triangle r/\rho_i\approx 0.2$.
We now compare this width with the size of the fine structures in the
electron distribution.

It is known that the fast motion of electrons along the equilibrium magnetic field lines leads to fine radial
structures in the linear eigenmodes of drift-waves. Consider far passing electrons.
The size of the fine structure is determined by the transition between
exact resonance at the rational surfaces of the $q$-profile and the adiabatic response far away from resonance.
The typical size of the fine structures has been estimated to be below
the ion Larmor radius. Following Dominski \cite{dominski2015}, we estimate the size of the fine structures according to
\beqn
\frac{\delta_{\rm lin}}{\rho_i}=2\frac{\omega_r}{v_{\rm Ti}/R}\frac{q_0}{\hat{s}\sqrt{m_i/m_e}k_y\rho_i},
\label{eq:delta}
\eeqn
where $\hat{s}\approx 0.8$ is the magnetic shear parameter. The estimated value is $\delta_{\rm lin}\approx 0.49\rho_i$. One sees
that, at $tc_s/a=5$, the size of the field line excursion is a significant fraction of the size of the fine structures in the
eigenmode.

We can now try to understand the strong effects of EPN. As the KBM grows linearly, the magnetic field line is
displaced in the radial direction. The radial displacement of the electrons due to magnetic fluttering reaches
the ion Larmor radius scale well before the $E\times B$ nonlinearity takes effect.
Once the radial excursion of the electrons is a significant fraction of 
 the size of the fine structures in the KBM eigenmode, the ballistic motion along the
field lines starts to turn the fine
structures of the distribution in radius into fine structures in velocity. This can be seen as follows.
In the absence of collisions, the total distribution function $f(x,y,z,\mu,v_\pa,t)$ is constant along the particle trajectories.
Let $\tau$ denote a time scale of the poloidal transit time of a thermal electron, $\tau\sim 2\pi qR/v_{\rm Te}$.
Consider the trajectory of a far passing electron with coordinates $(x,y,z,\mu,v_\pa)\equiv {\bf Z}$ at time $t$.
At the earlier time of $t-\tau$, the electron is located at ${\bf Z}-\triangle{\bf Z}$ with
$\triangle{\bf Z}\equiv[\triangle x(v_\pa,\tau),\triangle y(v_\pa,\tau),\triangle z(v_\pa,\tau),\triangle \mu,\triangle v_\pa(\tau)]$
(suppressing the dependence on other coordinates) and $\triangle \mu=0$.
As a result of magnetic fluttering
the displacements $\triangle x(v_\pa,\tau)$ and $\triangle y(v_\pa,\tau)$ 
vary with $v_\pa$. The distribution functions at $t$ and $t-\tau$ are related by, 
\beqn
f({\bf Z},t)=f({\bf Z}-\triangle{\bf Z},t-\tau),
\eeqn
and the velocity derivatives are related by
\beqn
\left.\frac{\partial f}{\partial v_\pa}\right|_{{\bf Z},t}
=\left.\frac{\partial f}{\partial v_\pa}\right|_{{\bf Z}-\triangle{\bf Z},t-\tau} 
-\frac{\partial \triangle x}{\partial v_\pa}
\left. \frac{\partial f}{\partial x}\right|_
{{\bf Z}-\triangle{\bf Z},t-\tau}
 +\ldots,
\label{eq:mixing}
 \eeqn
where on the right side
terms proportional to $\partial f/\partial y$ or $\partial f/\partial z$ are omitted as there are no fine structures
in $y$ and $z$. A correction to the first term involving $\partial \triangle v_\pa/\partial v_\pa$ is also omitted.
One sees that, over the period $\tau$, the velocity space gradient is produced in proportion to
the radial derivative of the distribution and the $v_\pa$-derivative of the radial displacement. If $\tau=t$  
then the first term on the right side stands for the contribution of the initial condition, which is small because
the initial distribution is smooth in velocity. If $\tau$ is small, then the contribution of the second term is negligible
because it is proportional to $\tau$. 
The process under discussion is the familiar phase-mixing of distribution function due to ballistic motion,
see Hammett \etal \cite{hammett1992} and Barnes \etal \cite{barnes2010} for discussions within the electrostatic gyrokinetic
model. Our focus here is the radial mixing due to the magnetic perturbations.

It is illustrative to estimate how fast 
the magnitude of the EPN term grows in time. This can be done by estimating the magnitude of the $\partial f/\partial x$ term
of Eq.~\ref{eq:mixing}. 
The time window is chosen to be $\tau=\pi q_0R_0/v_{\rm Te}$, or $\gamma_{\rm lin}\tau\approx 0.075$.
This choice is based on the consideration
that $\tau$ should be such as to allow a simple estimate of $\triangle x(v_\pa,\tau)$, but otherwise as large as possible,
so that the contribution of the $\partial f/\partial x$ term is significant.
Meanwhile, $\tau$ should be well below
the mode growth time ($1/\gamma_{\rm lin}$), so that the estimated EPN effect can be viewed as an instantaneous response.
Take $\partial f/\partial x\sim\df/\delta_{\rm lin}$.
The radial drift velocity due to magnetic fluttering
is $v_\pa\delta B_r/B_0$, hence $\triangle x\sim v_\pa\tau \delta B_r/B_0$, with $\delta B_r$ the magnitude of the average value over
the parallel distance $v_\pa\tau$ which, for a thermal electron with $v_\pa=v_{\rm Te}$,
is half of the field line length in a poloidal revolution. It is worth
noting that for a field-aligned mode such as KBM no severe cancellation occurs while taking the average. If $\tau$ is much larger
than the transit time, estimating $\triangle x$  would be more difficult. 
We now have the estimates $\partial \triangle x/\partial v_\pa\sim \tau \delta B_r/B_0$
and $\partial\df/\partial v_\pa\sim \tau(\delta B_r/B_0)(\df/\delta_{\rm lin})$.
For an estimate of
the nonlinear parallel acceleration we use for simplicity the inductive component
of the electric field, $E_\pa\sim\partial A_\pa/\partial t$, so that
$\dot{v}_{\pa 1}\sim (e/m_e)\partial A_\pa/\partial t\sim (e/m_e)\omega_r A_\pa$, $A_\pa$ being understood as the magnitude of
the average value.
Putting these estimates together we obtain an estimate of the newly generated EPN term,
\beqn
\dot{v}_{\pa 1}\frac{\partial \df}{\partial v_\pa}
\sim \frac{e}{m_e}\omega_r A_\pa \tau\frac{\delta B_r}{B_0}\frac{\df}{\delta_{\rm lin}}.
\eeqn
We now view the net effect of the EPN term as being equivalent to a damping term of the form $\gamma_{\rm nl}\df$.
The quantity $\gamma_{\rm nl}$ is a measure of how fast nonlinear damping is produced by EPN, per electron transit time.
That the EPN term in fact produces nonlinear damping is evidenced by the simulation. 
The equivalent damping rate is
\beqn
\gamma_{\rm nl}
\sim \frac{e}{m_e}\omega_r A_\pa \tau\frac{\delta B_r}{B_0}\frac{1}{\delta_{\rm lin}}.
\label{eq:gnl}
\eeqn
This damping rate can be compared to the KBM linear growth rate, which is $\gamma_{\rm lin}=0.34c_s/a$. Using the
magnetic perturbation data at $r/a\approx 0.5$ and $tc_s/a=5$, one obtains
$\gamma_{\rm nl}/\gamma_{\rm lin}=0.018$. Notice the quadratic dependence
of $\gamma_{\rm nl}$ on the amplitude of the magnetic perturbation.
The nonlinear damping increases rapidly in time as the mode grows. At $tc_s/a=10$ the amplitude $\abs{A_\pa}$ has increased to
$\sim 3$ times the level at $tc_s/a=5$, yielding $\gamma_{\rm nl}/\gamma_{\rm lin}\sim 0.16$. More importantly,
for the chosen $\tau$, the total velocity derivative
is likely dominated by the first term on the right side of Eq.~\ref{eq:mixing}.

The above reasoning is only an argument for the plausibility of EPN as a significant nonlinear mechanism.
A quantitative analysis of the nonlinear process requires the development of novel diagnostic tools for inspecting the distribution
function in a particle simulation, which we will pursue in the future. Nevertheless, the above reasoning already sheds
light on several numerical observations.

First, we note that
 the $E\times B$ drift is independent of the parallel velocity,
hence it does not provide a mechanism for generating velocity space structures as efficiently as the magnetic fluttering does. This
is consistent with the observation that EPN is not important for electrostatic instabilities.

\begin{figure}[!htp]
\includegraphics[width=0.6\textwidth]{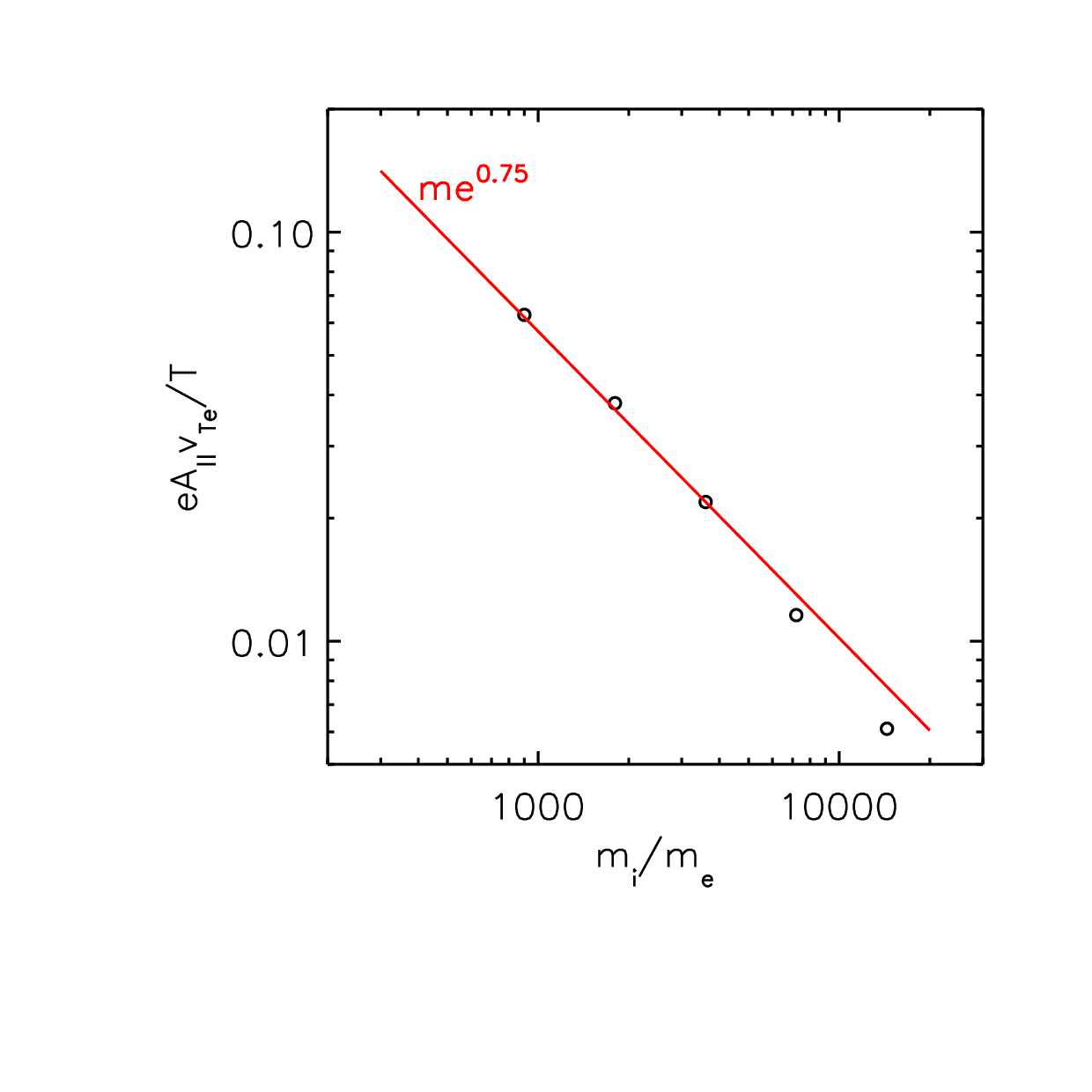}
\caption{Saturation amplitude of the volume-averaged amplitude, $eA_\pa v_{\rm Te}/T_0$, as a function of the
  ion-to-electron mass ratio. The solid line shows the scaling $A_\pa\propto m_e^{0.75}$.} 
\label{fig:apam}
\end{figure}

Second,
both the fine structure in radius and
the fast radial motion arise from the fast parallel streaming of the electrons, hence it
can be expected that the efficiency of EPN as a saturation mechanism is sensitive to the ion-to-electron mass ratio.
If we have chosen $\tau$, when estimating $\gamma_{\rm nl}$, to be the mode growth time $\tau=1/\gamma_{\rm lin}$
which is relatively insensitive to the mass ratio \cite{mishchenko2021}, then
the first term on the right side of Eq.~\ref{eq:mixing} can be neglected. We can still assume
$\partial \triangle x/\partial v_\pa \sim \tau\delta B^{\rm eff}/B_0$,  with an effective magnetic perturbation $\delta B^{\rm eff}$ 
that is proportional to $A_\pa(t)$, so that we can replace $\delta B_r$ in Eq.~\ref{eq:gnl}
with $\delta B^{\rm eff}=kA_\pa(t)$ with the constant $k$ independent of $m_e$.
Noticing from Eq.~\ref{eq:delta} that $\delta_{\rm lin}\propto \sqrt{m_e}$, Eq.~\ref{eq:gnl} then
indicates $\gamma_{\rm nl}\propto A_\pa^2 / m_e^{\rm 1.5}$.
If we now  assume $\gamma_{\rm nl}=\gamma_{\rm lin}$ at saturation,
the scaling $A_\pa\propto m_e^{0.75}$ is obtained for the saturation amplitude.
To test this scaling we have repeated 
the high-$\beta$ simulation of Fig.~\ref{fig:fluxcmpx} with electron masses ranging from $m_i/m_e=900$  to $m_i/m_e=14400$.
Fig.~\ref{fig:apam} shows the initial saturation amplitude of the vector potential
as a function of the mass ratio, 
normalized as $eA_\pa v_{\rm Te}/T_0$ with a fixed $v_{\rm Te}$ corresponding to $m_i/m_e=3600$.
The saturation amplitude scales with the electron mass approximately as $A_\pa \propto m_e^{0.75}$ for the low values of $m_i/m_e$,
and becomes more sensitive to $m_e$ as $m_e$ decreases.

Third, the importance of velocity space structures suggests that collisions could also be important.
To test this possibility,
the $\beta=0.04$ simulation of Fig.~\ref{fig:fluxcmpx} is also repeated with a much-reduced collision rate of $\nu_ea/c_s=0.036$.
The initial saturation level of the ion heat flux
is about the same as with $\nu a/c_s=0.36$. However, the initial saturation is followed by a period of slow
growth that leads to a final saturated flux of $Q_i/Q_{\rm GB}\approx 10$, much higher than the result in Fig.~\ref{fig:fluxcmpx}.

Although the discussion so far has focused on KBM, the radial excursion of the field lines and
the fine structures in the linear eigenmode are not peculiar to KBM. EPN is effective as long as $\beta$ is not too small.
We have carried out simulations over a broad range of beta for $\beta \geq 1.2\%$, and observed
strong EPN effects.

\begin{figure}[!htp]
\includegraphics[width=0.6\textwidth]{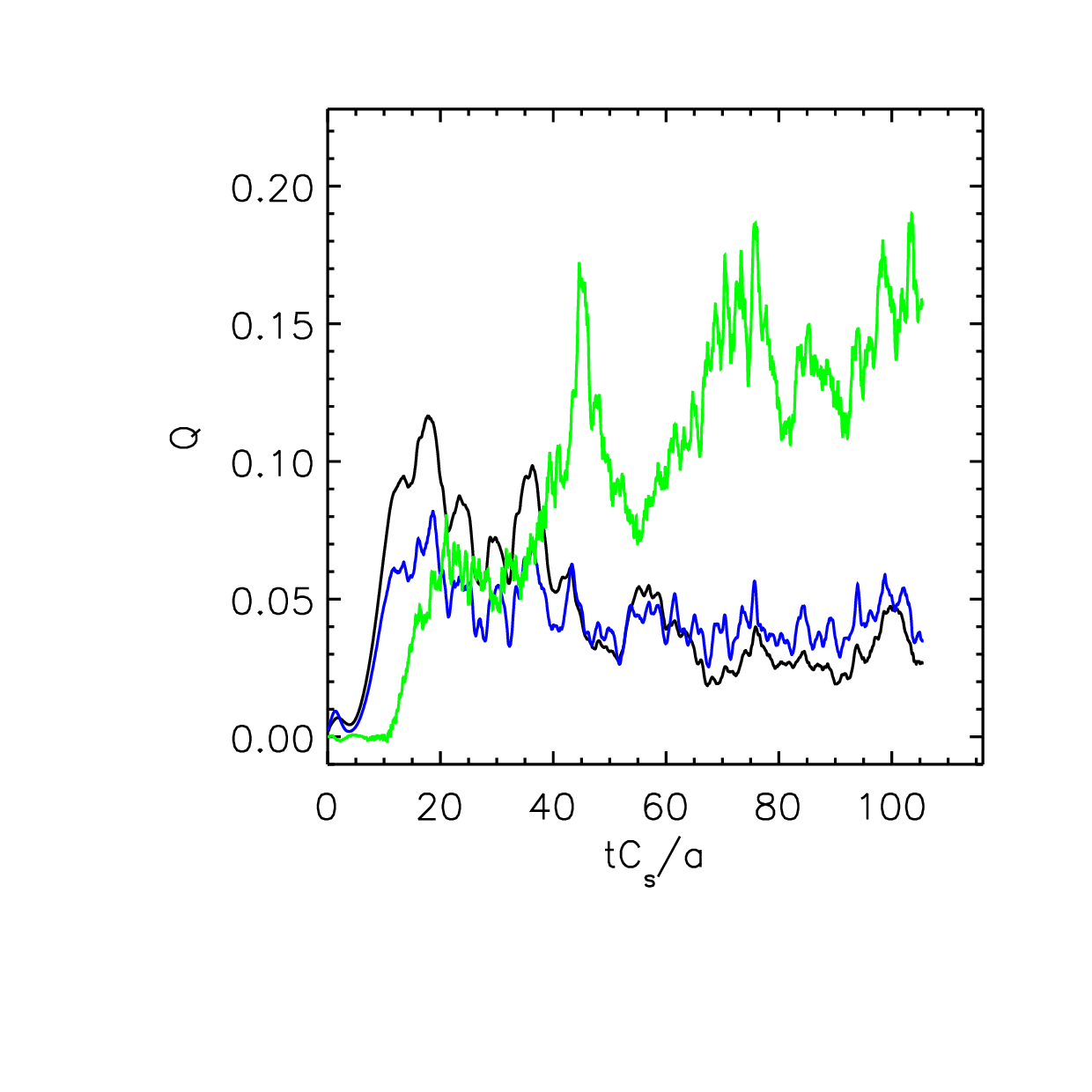} 
\caption{Time history of heat fluxes with multiple unstable modes with increased particle number and radial resolution. Black:
ion heat flux from $E\times B$; blue: electron heat flux from $E\times B$; green: electron heat flux from magnetic fluttering. }
\label{fig:fluxtmp}
\end{figure}

\begin{figure}[!htp]
\includegraphics[width=0.6\textwidth]{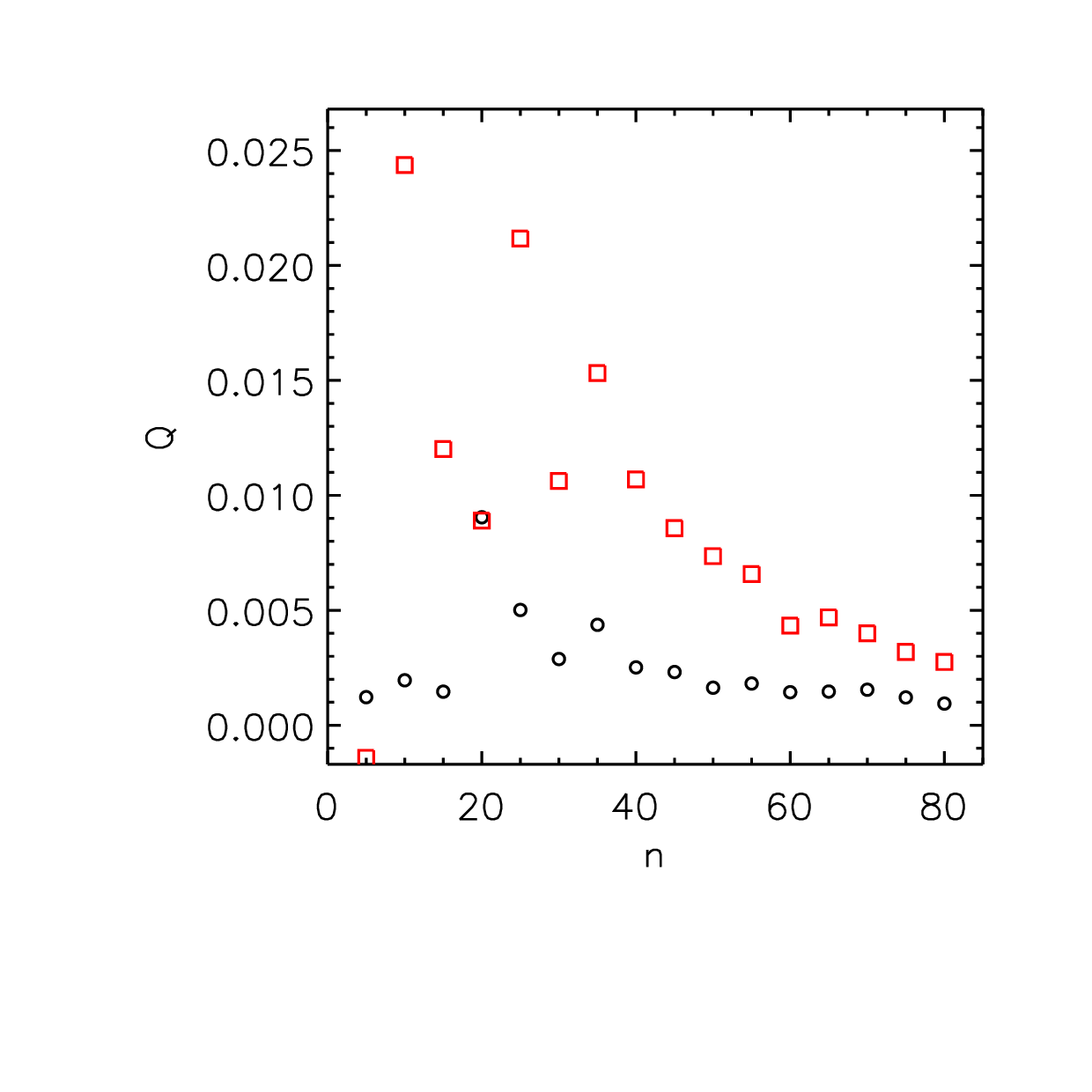} 
\caption{Heat fluxes vs. toroidal mode number. Circle: electron heat flux from $E\times B$; square: electron heat flux from
magnetic fluttering.}
\label{fig:fluxky}
\end{figure}

Finally, we present a simulation of $\beta=0.04$ to show the EPN effects in a conventional numerical setting,
i.e. with multiple unstable toroidal modes. The included mode numbers are $n=0,5,10,\ldots,80$. The binormal
grid numbers is doubled,  $n_y=128$. The particle number per grid is also doubled for each species. The time step is
$\triangle t c_s/a=0.00139$. The evolution of
heat fluxes is shown in Fig.~\ref{fig:fluxtmp}. The $E\times B$ heat fluxes reach a steady state,
but the electron heat flux due to magnetic
fluttering appears to grow slowly in time. We attribute this slow growth to the
slow cascading of the magnetic fluctuations to the high-$n$ toroidal modes, say $40\leq n \leq 80$. Fig.~\ref{fig:fluxky} shows 
the contributions to the electron heat flux from
individual toroidal modes, averaged over the second half of the time history. 
the magnetic contribution has a broad spectrum,
indicating the cascading of the magnetic energy to high-$n$ modes.

\newpage
\section*{Acknowledgement}
 This research was supported by the U.S. Department of
 Energy's SciDAC project ABOUND: Developing multiscale simulation of boundary plasma dynamics under grant SCW1832, 
 the SciDAC project FRONTIERS in Leadership Gyrokinetic Simulation under grant DE-SC0024425.
This research used resources of the National Energy Research
Scientific Computing Center, which is supported by the Office of
Science of the U.S. Department of Energy under Contract No. 
DE-AC02-05CH11231.
\section*{Conflict of interest}
The authors have no conflicts to disclose.
\section*{Data availability}
        The data that support the findings of this study are available from the corresponding author upon reasonable request.


\end{document}